\documentclass[aps,pra,twocolumn,showpacs]{revtex4-1}
\usepackage{ulem}
\usepackage{color, soul} 
\usepackage{indentfirst}
\usepackage{bm}
\usepackage{verbatim}
\usepackage{graphicx}
\usepackage{subfigure}
\usepackage{epsfig}
\usepackage{amsmath}
\usepackage{booktabs} 
\usepackage{multirow}
\usepackage{txfonts} 
 \usepackage{float}
 \usepackage{tabularx}
 \usepackage{threeparttable}

\usepackage[citecolor=blue,colorlinks=true,linkcolor=blue,urlcolor=blue,breaklinks=true]{hyperref}

\begin{document}

\title{Cyclotron dynamics of a Bose-Einstein condensate in a quadruple-well potential with synthetic gauge fields }%
\author{Wen-Yuan Wang$^{1,2}$}
\author{Ji Lin$^{3}$}   
\author{Jie Liu$^{4,5}$}
\email[E-mail address: ]{jliu@gscaep.ac.cn.}
\affiliation{$^{1}$Beijing Computational Science Research Center, Beijing 100193, China \\
$^{2}$Key Laboratory of Atomic and Molecular Physics $\&$ Functional Materials of Gansu Province, College of Physics and Electronic Engineering, Northwest Normal University, Lanzhou 730070, China\\
$^{3}$Department of Physics, Zhejiang Normal University, Jinhua 321004, China\\
$^{4}$Graduate School of China Academy of Engineering Physics, Beijing 100193, China\\
$^{5}$HEDPS, Center for Applied Physics and Technology, and College of Engineering, Peking University, Beijing 100871, China
}
\begin{abstract}
We investigate the cyclotron dynamics of Bose-Einstein condensate (BEC) in a quadruple-well potential with synthetic gauge fields. We use laser-assisted tunneling to generate large tunable effective magnetic fields for BEC. The mean position of BEC follows an orbit that simulated the cyclotron orbits of charged particles in a magnetic field. In the absence of atomic interaction, atom dynamics may exhibit periodic or quasi-periodic cyclotron orbits. In the presence of atomic interaction, the system may exhibit self-trapping, which depends on synthetic gauge fields and atomic interaction strength. In particular, the competition between synthetic gauge fields and atomic interaction leads to the generation of several discontinuous parameter windows for the transition to self-trapping, which is obviously different from that without synthetic gauge fields.
\end{abstract}
 \maketitle
\section{Introduction}
The Bose-Einstein condensate (BEC) constitutes a unique platform to explore new physical regimes in condensed matter systems due to the remarkable feature of a high degree controlled environment \cite{RevModPhys.80.885}. The interplay between magnetic fields and interacting charged particles can exhibit seminal quantum many-body phenomena, such as topological insulators \cite{RevModPhys.82.3045, Qian_Niu:43601,Yang2015}, the integer \cite{PhysRevLett.45.494}, and the fractional \cite{PhysRevLett.48.1559,PhysRevLett.50.1395} quantum Hall effects. BEC can exploit effective synthetic gauge fields by implementing complex hopping amplitudes characterized by a Peierls phase \cite{Jaksch_2003,Goldman_2014,NatRevPhys.2.8,Zhang31}, which has been achieved in experiments \cite{RevModPhys.81.647,RevModPhys.83.1523,RevModPhys.86.153,RevModPhys.91.015005}. Several schemes have been proposed to realize synthetic gauge fields, such as using rotating optical lattices \cite{PhysRevLett.84.806,Abo-Shaeer476,PhysRevLett.97.240402,PhysRevLett.104.050404,RevModPhys.81.647,doi:10.1080/00018730802564122}, laser-assisted tunneling in an optical superlattice \cite{PhysRevLett.107.255301,PhysRevLett.111.185302,PhysRevLett.111.185301,2013Experimental}, implementing synthetic dimensions \cite{Mancini1510,Stuhl1514,RevModPhys.87.637}, periodic driving of the optical lattice \cite{Kolovsky_2011,PhysRevLett.108.225304,Jotzu2014}, and extension to engineer density-dependent gauge fields \cite{PhysRevLett.121.030402,NatPhys.15.1168,NatPhys.15.1161,PhysRevX.10.021031}. BEC with strong synthetic gauge fields has been realized by laser-assisted tunneling processes in a tilted lattice potential \cite{Kennedy2015}.

Cyclotron orbit is one of the typical dynamic characteristics of charged particles moving in a magnetic field. Since the neutral atoms can be operating as charged particles by engineering artificial gauge potentials, quantum cyclotron orbits of charged neutral ultracold atom have been observed experimentally \cite{PhysRevLett.107.255301,PhysRevLett.111.185301,2013Experimental}. Since the cyclotron orbital motion play an essential role in the emergence of several novel phenomena, it would be significant to further study the characteristics of cyclotron dynamics of ultracold atoms with synthetic gauge fields. In the current work, we propose a possible scheme to study cyclotron dynamics of a BEC in a quadruple-well potential with synthetic gauge fields. We are focused primarily on the significance of atomic interaction to cyclotron dynamics. In the absence of atomic interaction, we analytically provide the conditions for the occurrence of periodic and quasi-periodic cyclotron orbits. In the presence of atomic interaction, the periodicity of cyclotron dynamics disappears completely. With sufficiently strong interactions, cyclotron dynamics is completely suppressed, and the bosons form a dynamically localized state, analogous
to self-trapping effects observed in BEC in double-well potentials \cite{PhysRevA.55.4318,PhysRevLett.79.4950,PhysRevLett.95.010402,PhysRevA.74.063610}. The phase diagrams of transition to self-trapping are obtained, which are affected by synthetic gauge fields and atomic interaction strength. In particular, there are several discontinuous parameter windows for the transition to self-trapping, which is obviously different from the situation without synthetic gauge fields.

The paper is organized as follows. In section II, we introduce the model and Hamiltonian of a BEC in a quadruple-well potential with the presence of synthetic gauge fields. In section III, we study cyclotron dynamics both in the absence of atomic interaction and in the presence of atomic interaction. Finally, we present a summary and conclusion in section IV.

\section{Physical model and Hamiltonian}
\begin{figure}[tb]
\centering
\includegraphics[width=\columnwidth]{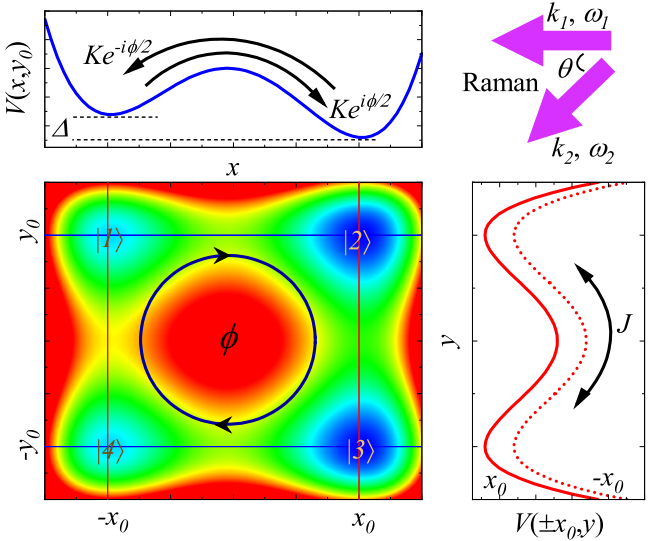}
\caption{(color online) Schematic diagram showing the system under consideration. The quadruple-well potential can be equivalently with a four-site plaquette in an optical lattice, which was achieved by applying superlattice potentials along both the $x$- and $y$-directions, so that all dynamics is restricted to four sites without any coupling between plaquettes. The potential along the $y$-direction is symmetric, and the bare tunneling occurs along the $y$-direction with amplitude $J$. The potential along the $x$-direction generates a tilted potential with amplitude $\Delta$, which inhibits the bare tunneling along the $x$-direction. Then, an additional pair of lasers with wave vectors $\textbf{k}_{1,2}$ and frequency difference $\omega_1-\omega_2=\Delta/\hbar$ induced resonant tunneling along the $x$-direction with complex amplitude $K(\textbf{R})$ whose phase depends on position.}
\label{fig:model}
\end{figure}

We consider a BEC trapped in a two-dimensional quadruple-well potential. The quadruple-well potential can be created by the superposition of two sets of double-well potentials \cite{PhysRevLett.106.025302} along with both the $x$- and $y$- directions. For simplicity, we assume the double-well potential along the $y$-direction to be symmetric as shown in Fig. \ref{fig:model}, with each well having the same harmonic trapping frequency $\omega_y$. The double-well potential along the $x$-direction generates a tilted potential with amplitude $\Delta$. The four wells of the quadruple-well are denoted as 1, 2, 3, 4 as shown in Fig. \ref{fig:model}. Such a quadruple-well potential created by superposition double-well potential can be generated in experiments \cite{PhysRevLett.106.025302}, with the form
\begin{eqnarray}
V(x, y)=V_{0x}(x^2-x^2_0)^2+V_{0y}(y^2-y^2_0)^2-\frac{\Delta}{2}x,
\label{eq:potentials}
\end{eqnarray}
where the parameters $V_{0x}$ and $V_{0y}$ are both tunable in the experiments  \cite{PhysRevLett.106.025302}. Expanding the first and second terms in $V(x, y)$ near $\pm x_0$ and $\pm y_0$ in $x$ and $y$ directions respectively, one obtains the harmonic form as $V(\pm x)=\frac{1}{2}m\omega^2_x (x\pm x_0)^2$, $V(\pm y)=\frac{1}{2}m\omega^2_y (y\pm y_0)^2$, with $m$ indicating the mass of an atom. Thus $V_{0x}=m\omega^2_x/8x_0^2$ and $V_{0y}=m\omega^2_y/8y_0^2$. Besides, in the third term in $V(x, y)$, a magnetic field gradient along the $x$-direction is used to generate the energy offset of $\Delta$ between two wells along the $x$-direction, as shown in Fig. \ref{fig:model}, and normal tunneling is inhibited along the $x$-direction due to the energy offset of $\Delta$. A pair of Raman lasers with frequency difference $\omega=\omega_1-\omega_2=\Delta/\hbar$ induce resonant tunneling along the $x$-direction while hopping along the $y$-direction is controlled by the depth of the potential along this direction.

The system can be described by the following Hamiltonian in the interaction picture,
\begin{eqnarray}\label{ha}
\hat{H}&=&-K(e^{i\phi/2}\hat{a}^{\dagger}_{1}\hat{a}_{2}+e^{i\phi/2}\hat{a}^{\dagger}_{3}\hat{a}_{4}+H.c.)
                   -J(\hat{a}^{\dagger}_{2}\hat{a}_{3}+\hat{a}^{\dagger}_{4}\hat{a}_{1}+H.c.)\nonumber\\
             &&+\frac{U}{2}\sum_{j}\hat{n}_{j}(\hat{n}_{j}-1).
\end{eqnarray}
Here the operator $\hat{a}_{j}~(\hat{a}_{j}^{\dag})$ is the bosonic annihilation (creation) operator,  $\hat{n}_{j}=\hat{a}_{j}^{\dag} \hat{a}_{j}$ is the local number operator on the $j$-well. $J$ is the regular tunneling term, $Ke^{\pm i\phi/2}$ is the Raman laser induced tunneling term, and $U$ is the on-site interaction strength with positive (negative) values denoting repulsive (attractive) interaction.

When $N\rightarrow \infty$, the semiclassical limit of the second-quantized Hamiltonian can be achieved by the mean-field model. Then, one can replace the annihilation and creation operators by their respective expectation values with complex numbers,
\begin{eqnarray}\label{expectation}
\hat{a}_{j}\simeq \langle\hat{a}_{j}\rangle\equiv\psi_j, ~~~\hat{a}^{\dagger}_{j}\simeq\langle\hat{a}^{\dagger}_{j}\rangle\equiv\psi_j^*.
\end{eqnarray}
Since the complex numbers commute in contrast to the quantum mechanical operators, we will begin on the many particle side with the mean-field model Hamiltonian in the following. Note that the mean-field wave function is normalized as $\sum_j|\psi_j|^2=1$.

Then, we obtain the energy functional of the Hamiltonian (\ref{ha}) as given by
\begin{eqnarray}\label{e}
E&=&-K(e^{i\phi/2}\psi_{1}^{*}\psi_{2}+e^{i\phi/2}\psi_{3}^{*}\psi_{4}+c.c.)
                   -J(\psi_{2}^{*}\psi_{3}+\psi_{4}^{*}\psi_{1}+c.c.)\nonumber\\
             &&+\frac{U}{2}\sum_{j}|\psi_{j}|^4.
\end{eqnarray}
The time evolution of the complex-valued mean-field amplitudes can be obtained by $i\frac{\partial \psi_{j}}{\partial t}=\frac{\partial E}{\partial \psi^{*}_{j}}$, which gives the following coupled Gross-Pitaevskii equations,
\begin{eqnarray}\label{gp}
&&i\frac{\partial \psi_{1}}{\partial t}=-Ke^{i\phi/2}\psi_{2}-J\psi_{4}+U|\psi_{1}|^2\psi_{1}, \nonumber\\
&&i\frac{\partial \psi_{2}}{\partial t}=-Ke^{-i\phi/2}\psi_{1}-J\psi_{3}+U|\psi_{2}|^2\psi_{2}, \nonumber\\
&&i\frac{\partial \psi_{3}}{\partial t}=-Ke^{i\phi/2}\psi_{4}-J\psi_{2}+U|\psi_{3}|^2\psi_{3}, \nonumber\\
&&i\frac{\partial \psi_{4}}{\partial t}=-Ke^{-i\phi/2}\psi_{3}-J\psi_{1}+U|\psi_{4}|^2\psi_{4}.
\end{eqnarray}

It is worth emphasizing that the quadruple-well potential can also be realized equivalently with a four-site plaquette in an optical lattice in experiments \cite{PhysRevLett.107.255301,PhysRevLett.111.185301,2013Experimental}. Following the experimental realization of strong tunable effective magnetic fields in an optical superlattice \cite{PhysRevLett.107.255301,PhysRevLett.111.185301,2013Experimental}, the four-site plaquette was isolated in a two-dimensional optical lattice, which was achieved by applying superlattice potentials along both the $x$- and $y$-directions, so that all dynamics is restricted to four sites without any coupling between plaquettes \cite{PhysRevLett.107.255301,PhysRevLett.111.185301,2013Experimental}.

\section{Cyclotron dynamics}
We now study the cyclotron dynamics of a BEC in a quadruple-well potential to exhibit the influences of the synthetic gauge field and atomic interaction on the particle
flow.
\subsection{Cyclotron dynamics in the absence of atomic interaction}
\subsubsection{Phase diagram of periodic and quasi-periodic cyclotron orbits}
We first consider the strong coupling-tunneling regime, so that the atomic interaction term is neglected, i.e., $U=0$. For this linear case, the system is analytically solvable. The
solutions of ${\psi_{1}, \psi_{2}, \psi_{3}, \psi_{4}}$ are determined by the initial conditions. For the initial conditions that we are concerned with in this paper, i.e., ${ \psi_{1}(0)=\sqrt{0.5}, \psi_{2}(0)=0, \psi_{3}(0)=0, \psi_{4}(0)=-\sqrt{0.5} }$, they are given by
\begin{eqnarray}\label{dyn}
&&\psi_{1}=\frac{1}{2\sqrt{2}}\cos\alpha t +\frac{1}{2\sqrt{2}}\cos\beta t +c_1\sin\alpha t +c_2\sin\beta t, \nonumber\\
&&\psi_{2}=-\frac{1}{2\sqrt{2}}\cos\alpha t +\frac{1}{2\sqrt{2}}\cos\beta t +c_1\sin\alpha t -c_2\sin\beta t, \nonumber\\
&&\psi_{3}=\frac{1}{2\sqrt{2}}\cos\alpha t -\frac{1}{2\sqrt{2}}\cos\beta t +c_1\sin\alpha t -c_2\sin\beta t, \nonumber\\
&&\psi_{4}=-\frac{1}{2\sqrt{2}}\cos\alpha t -\frac{1}{2\sqrt{2}}\cos\beta t +c_1\sin\alpha t +c_2\sin\beta t.
\end{eqnarray}
Here, $c_1=i(J+K\exp(i\phi/2))/(2\sqrt{2}\alpha)$, $c_2=i(J-K\exp(i\phi/2))/(2\sqrt{2}\beta)$, $\alpha=\sqrt{K^2+J^2+2KJ\cos{(\phi/2)}}$, and $\beta=\sqrt{K^2+J^2-2KJ\cos{(\phi/2)}}$.

Cyclotron orbits of the average particle position obtained from the mean atom positions,
\begin{eqnarray}\label{centre}
&&x_{c}=\langle x\rangle/x_0=-|\psi_{1}|^2+|\psi_{2}|^2+|\psi_{3}|^2-|\psi_{4}|^2, \nonumber\\
&&y_{c}=\langle y\rangle/y_0=|\psi_{1}|^2+|\psi_{2}|^2-|\psi_{3}|^2-|\psi_{4}|^2.
\end{eqnarray}
\begin{figure}[tb]
\centering
\includegraphics[width=\columnwidth]{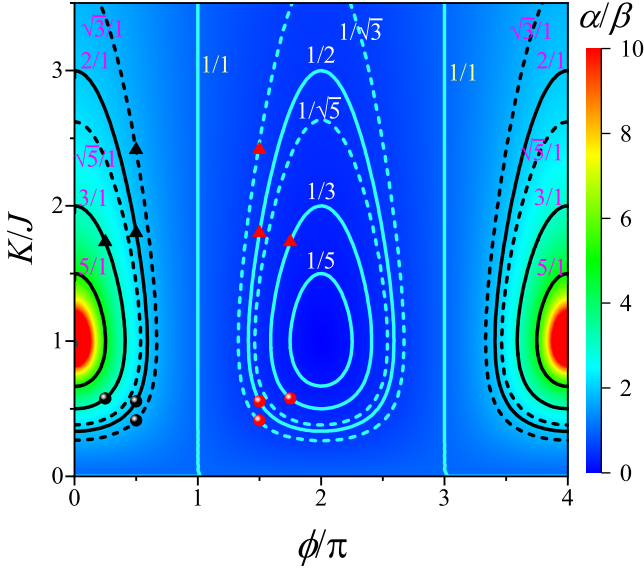}
\caption{(color online) Contour plot of the rate $\alpha/\beta$ as a function of synthetic gauge fields $\phi/\pi$ and rescaled tunneling amplitude $K/J$. Periodic and quasi-periodic dynamic cyclotron orbits can be analyzed by the principle of commensurability between $\alpha$ and $\beta$. When these two frequencies $\alpha$ and $\beta$ are commensurable, the dynamics of the system presents periodic cyclotron orbits. However, when the two frequencies are incommensurable, the dynamics of the system presents quasi-periodic cyclotron orbits. Typical frequency ratios are shown in the figure. }
\label{fig:rate}
\end{figure}
From equations (\ref{dyn}) and (\ref{centre}), one can obtain
\begin{eqnarray}\label{centre1}
&&x_{c}=\frac{K^2-J^2}{\alpha\beta}\sin\alpha t\sin\beta t-\cos\alpha t\cos\beta t, \nonumber\\
&&y_{c}=\frac{K\sin{(\phi/2)}}{\beta}\cos\alpha t\sin\beta t-\frac{K\sin{(\phi/2)}}{\alpha}\cos\beta t\sin\alpha t.
\end{eqnarray}
The equations  (\ref{centre1}) show that particle flow is significantly affected by synthetic gauge field.

It can be determined from equation (\ref{centre1})  that the dynamic cyclotron orbit significantly depends on the values of the two frequencies $\alpha$ and $\beta$. Periodic and quasi-periodic dynamic cyclotron orbit can be analyzed by the principle of commensurability between $\alpha$ and $\beta$. When these two frequencies $\alpha$ and $\beta$ are commensurable, the dynamics of the system presents periodic cyclotron orbits. However, when the two frequencies are incommensurable, the dynamics of the system presents quasi-periodic cyclotron orbits.

In Fig. \ref{fig:rate}, we show the rate $\alpha/\beta$ as function of synthetic gauge fields $\phi/\pi$ and rescaled tunneling amplitude $K/J$. In this parameter space, the commensurable density of the two frequencies $\alpha$ and $\beta$ is related to both the synthetic gauge fields and rescaled tunneling amplitude. This ratio $\alpha/\beta$ shows a periodic change to synthetic gauge fields, and the period is $4\pi$.  In a period, as shown in Fig. \ref{fig:rate}, the ratio is symmetric about $\phi=2\pi$. When $\phi$ is equal to an odd multiple of $\pi$, $\alpha/\beta\equiv1$ no matter how the rescaled tunneling amplitude $K/J$ changes. It means that the dynamics of the system always present a periodic cyclotron orbit. Typical frequency ratios are shown in Fig. \ref{fig:rate}. For a certain synthetic gauge field $\phi$, one can always find a series of tunneling amplitude $K/J$, such that the system exhibit a periodic cyclotron orbit.

It is worth noting that although sometimes the ultracold atoms in the artificial gauge field might be similar to the electrons in the magnetic field, their physical schemes are quite different. For example, considering electrons moving in a two-dimensional square lattice, the corresponding Hamiltonian can be described by the celebrated Harper Hamiltonian \cite{ProcPhysSocA.68.874,PhysRevB.14.2239}. If we also consider the motion of an electron in the four site square plaquette lattice under a uniform magnetic field, the cyclotron motion of an electron may be periodic or quasi-periodic, depending on the ratio of parameters $\tilde{\alpha}=\tilde{J}\sqrt{2+2\cos{(\phi/2)}}$ and $\tilde{\beta}=\tilde{J}\sqrt{2-2\cos{(\phi/2)}}$. Here, $\tilde{J}$ is the tunnel energy to nearest neighbors in the absence of a magnetic field. In both systems,  whether their motions are  periodic or quasi-periodic depend on the commensurability of these two parameters. However, for an electron in  a uniform magnetic field, both the parameters $\tilde{\alpha}$ and $\tilde{\beta}$ depend on $\tilde{J}$ and $\phi$. While,  for the ultracold atoms under artificial gauge field, the parameters $\alpha$ and $\beta$ depend on both  $J$ and  $K$ as well as  $\phi$.  With considering the  additional parameter $K$  and changing the ratio of $K/J$, one  can manipulate the  cyclotron motion and observe the phase transitions  according to our  phase diagram of Fig. \ref{fig:rate}.
\begin{figure}[tb]
\centering
\includegraphics[width=\columnwidth]{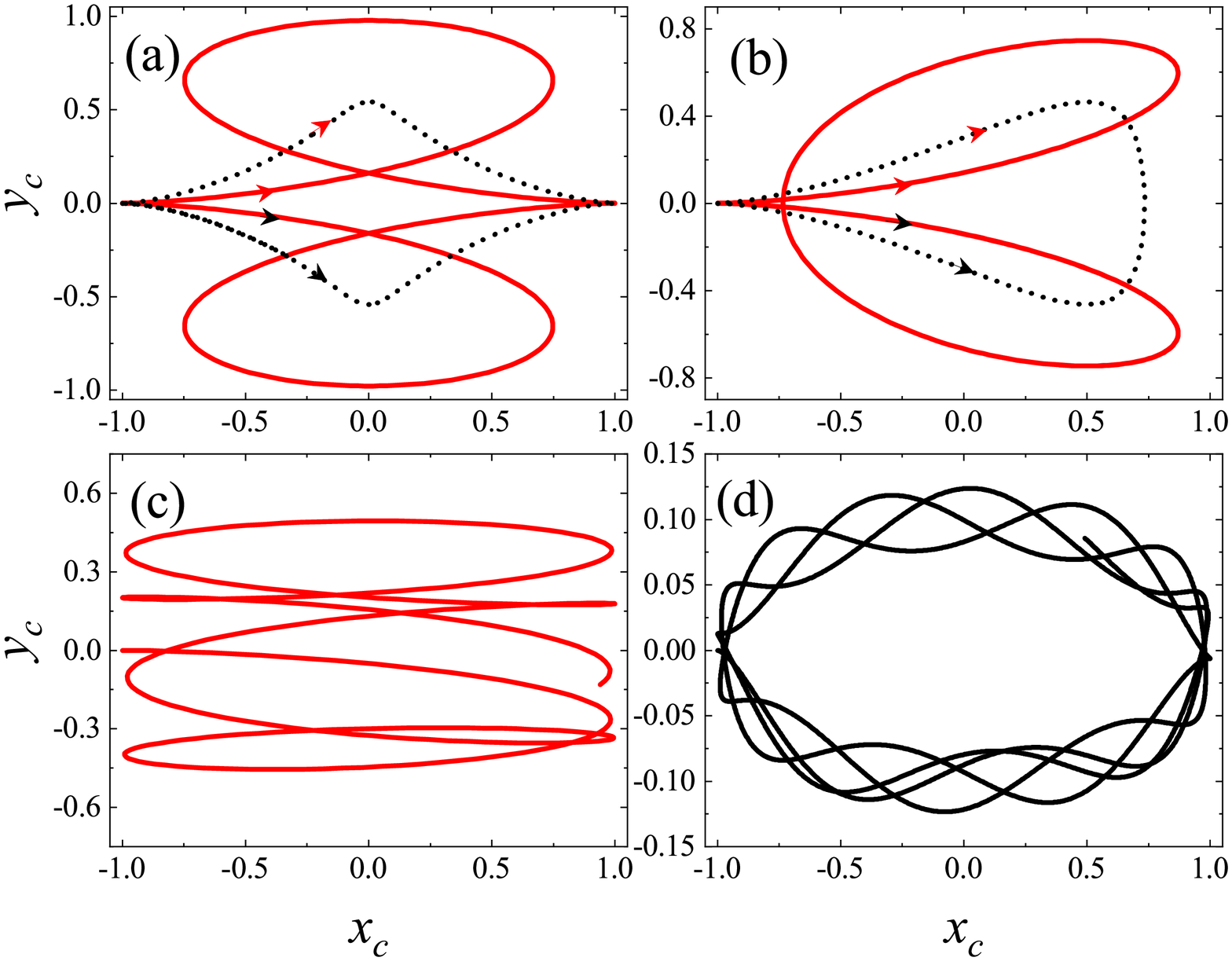}
\caption{(color online) The typical periodic and quasi-periodic orbits, the direction of the arrow represents the evolution direction of the orbit. (a). Periodic orbits for $\alpha/\beta=1/2$ and $\alpha/\beta=2/1$. (b). Periodic orbits for $\alpha/\beta=1/3$ and $\alpha/\beta=3/1$. In (a), the red solid line with the black arrow corresponds to the black triangle on the $\alpha/\beta=2/1$ line in Fig. \ref{fig:rate}; the black dashed line with the black arrow corresponds to the black point on the $\alpha/\beta=2/1$ line in Fig. \ref{fig:rate}; the red solid line with the red arrow corresponds to the red triangle on the $\alpha/\beta=1/2$ line in Fig. \ref{fig:rate}; the black dashed line with the red arrow corresponds to the red point on the $\alpha/\beta=1/2$ line in Fig. \ref{fig:rate}. (b) has an analogous conrrespondence for $\alpha/\beta=3/1, 1/3$. (c, d). Quasi-periodic orbits for $\alpha/\beta=\sqrt{3}/1$ at $\phi=0.5\pi$. (c) for the quasi-periodic cyclotron orbit corresponds to the black triangle on the $\alpha/\beta=\sqrt{3}/1$ line in Fig. \ref{fig:rate}. (d) for the quasi-periodic cyclotron orbit corresponds to the black point on the $\alpha/\beta=\sqrt{3}/1$ line in Fig. \ref{fig:rate}. }
\label{fig:orbits}
\end{figure}

In Fig. \ref{fig:orbits}, we show the periodic and quasi-periodic orbits for the typical frequencies $\alpha$ and $\beta$ as shown in Fig. \ref{fig:rate}.
For the rate $\alpha/\beta=2/1$ at $\phi=0.5\pi$, there are two sets of $K/J$ values (one is marked as the black triangle and another is marked as the black point in Fig. \ref{fig:rate}). The dynamics of the system indeed present a periodic cyclotron orbit at these two sets of $K/J$ values (shown in Fig. \ref{fig:orbits}(a)). However, the two periodic cyclotron orbits are obviously different. In \ref{fig:orbits}(a), the periodic cyclotron orbit of the red solid line with the black arrow corresponds to the black triangle on the $\alpha/\beta=2/1$ line in Fig. \ref{fig:rate}. The periodic cyclotron orbit of the black dashed line with the black arrow corresponds to the black point on the $\alpha/\beta=2/1$ line in Fig. \ref{fig:rate}. For the rate $\alpha/\beta=1/2$ at $\phi=1.5\pi$, there are also two values of $K/J$ (one is marked as the red triangle and another is marked as a red point in Fig. \ref{fig:rate}). In \ref{fig:orbits}(a), the red solid line with the red arrow corresponds to the red triangle on the $\alpha/\beta=1/2$ line in Fig. \ref{fig:rate}; the black dashed line with the red arrow corresponds to the red point on the $\alpha/\beta=1/2$ line in Fig. \ref{fig:rate}. It is very interesting that for different values of $\alpha/\beta=2/1$ and $\alpha/\beta=1/2$, the periodic cyclotron orbits can be the same, but their directions are opposite. In fact, these results can be obtained by analyzing the symmetry of Eq. (\ref{centre1}). One can get the following relationship, $x_c(\alpha, \beta)=x_c(\beta, \alpha)$ and $y_c(\alpha, \beta)=y_c(\beta, \alpha)$. Therefore, for the same value of $K/J$, the cyclotron orbit should be the same for $\alpha/\beta=2/1$ and $\alpha/\beta=1/2$. Due to the synthetic gauge fields difference $\pi$ between $\alpha/\beta=1/2$ and $\alpha/\beta=2/1$, the directions of the periodic cyclotron orbits in the $y$-direction are opposite (shown in Fig. \ref{fig:orbits}(a)).

\begin{figure}[tb]
\centering
\includegraphics[width=\columnwidth]{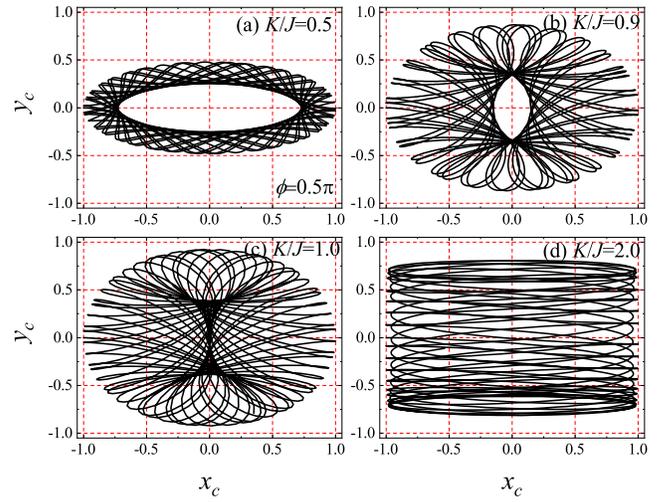}
\caption{(color online) Various quasi-periodic cyclotron orbits induced by rescaled tunneling amplitude $K/J$. The surrounding properties of quasi-periodic cyclotron orbits are significantly affected by tunneling amplitude.}
\label{fig:orbit1}
\end{figure}

We also show the periodic cyclotron orbit for the rates of frequency $\alpha/\beta=3/1$ and $\alpha/\beta=1/3$ in Fig. \ref{fig:orbits}(b)). Compared with the case of $\alpha/\beta=2/1$ or $\alpha/\beta=1/2$, the periodic cyclotron orbits in the cases $\alpha/\beta=3/1$ or $\alpha/\beta=1/3$ are completely different. In \ref{fig:orbits}(b), the periodic cyclotron orbit of the red solid line with the black arrow corresponds to the black triangle on the $\alpha/\beta=3/1$ line in Fig. \ref{fig:rate}. The periodic cyclotron orbit of the black dashed line with the black arrow corresponds to the black point on the $\alpha/\beta=3/1$ line in Fig. \ref{fig:rate}. The red solid line with the red arrow corresponds to the red triangle on the $\alpha/\beta=1/3$ line in Fig. \ref{fig:rate}. The black dashed line with the red arrow corresponds to the red point on the $\alpha/\beta=1/3$ line in Fig. \ref{fig:rate}. In fact, a large number of numerical results exhibit that although the dynamics of the system can show periodic orbits when the two frequencies are commensurate, the shapes of the orbits are quite different for different of frequency ratios $\alpha/\beta$.

Periodic and quasi-periodic dynamic cyclotron orbits can be analyzed by the principle of commensurability between $\alpha$ and $\beta$. When the two frequencies are incommensurable, the dynamics of the system presents a quasi-periodic cyclotron orbit. In Fig. \ref{fig:orbits}(c, d), we show the typical quasi-periodic orbits for $\alpha/\beta=\sqrt{3}/1$ at $\phi=0.5\pi$. Fig. \ref{fig:orbits}(c) for the quasi-periodic cyclotron orbit corresponds to the black triangle on the $\alpha/\beta=\sqrt{3}/1$ line in Fig. \ref{fig:rate}. Fig. \ref{fig:orbits}(d) for the quasi-periodic cyclotron orbit corresponds to the black point on the $\alpha/\beta=\sqrt{3}/1$ line in Fig. \ref{fig:rate}. These results further confirm our analysis of quasi-periodic cyclotron orbits.

\subsubsection{Various quasi-periodic cyclotron orbits}
\begin{figure}[tb]
\centering
\includegraphics[width=\columnwidth]{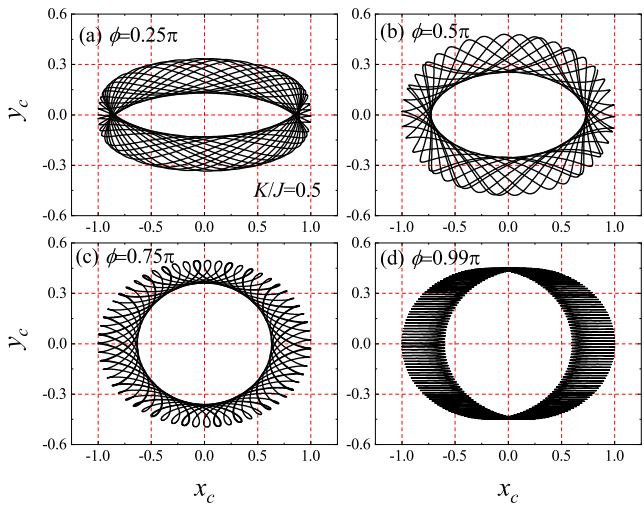}
\caption{(color online) Various quasi-periodic cyclotron orbits induced by synthetic gauge fields $\phi$. The surrounding properties of quasi-periodic cyclotron orbits cannot be changed by synthetic gauge fields. The volume of the phase space is affected by the synthetic gauge fields $\phi$.}
\label{fig:orbit2}
\end{figure}
Due to the presence of the synthetic gauge fields, the mean atom position follows an orbit that is a quantum analog of the cyclotron orbits for charged particles. This behavior is reminiscent of the Lorentz force acting on a charged particle in a magnetic field. Now, we carefully study the influences of both the rescaled tunneling amplitude $K/J$ and synthetic gauge fields $\phi$ on the quasi-periodic cyclotron dynamics.

In Fig. \ref{fig:orbit1}, we show various quasi-periodic cyclotron orbits induced by rescaled tunneling amplitude $K/J$ at $\phi=0.5\pi$. The surrounding properties of quasi-periodic cyclotron orbits can be varied by the tunneling amplitude. There is a critical rescaled tunneling amplitude $K_c/J=1$. When $K<K_c$, the mean atom position presents the trajectory of an elliptical ring solenoid, which means that it never reaches balance in $x$ and in $y$ directions at the same time. Meanwhile, the volume of the mean atom position in the phase space expands as the $K/J$ increases (as shown in Fig. \ref{fig:orbit1}(a, b)). When $K>K_c$, the mean atom position exhibits the trajectory of a cylindrical solenoid, and the volume of the mean atom position in the $y$-direction decreases as the $K/J$ increases (as shown in Fig. \ref{fig:orbit1}(d)). When the value of $K/J$ is large enough, then the volume of the mean atom position in the phase space along the $y$-direction would be suppressed.

The effect of synthetic gauge fields $\phi$ on the quasi-periodic cyclotron orbit is shown in Fig. \ref{fig:orbit2}. Compared with the effect of rescaled tunneling amplitude $K/J$, the surrounding properties of quasi-periodic cyclotron orbit cannot be changed by synthetic gauge fields $\phi$.

The study of the cyclotron motion of neutral ultracold atoms has received great interest, such as the Gaussian wave packet of the noninteracting atoms in optical lattices subjected to an additional harmonic trap potential \cite{Zhang_2020}. The authors shown that the harmonic trap potential plays a key role in modifying the equilibrium state properties of the system and stabilizing the cyclotron orbits of the condensate. Using the Gaussian ansatz which defines a Gaussian distribution of the atoms centered at the optical lattices and assuming the widths of the condensate unchanged with time. The authors obtained approximately cyclotron orbits under the condition of small cyclotron radius. Both physical model and mathematic treatment are completely different from ours even though the authors also study the periodic and quasi-periodic cyclotron motions.

\subsection{Cyclotron dynamics in the presence of atomic interaction: transition to self-trapping}
\begin{figure}[tb]
\centering
\includegraphics[width=\columnwidth]{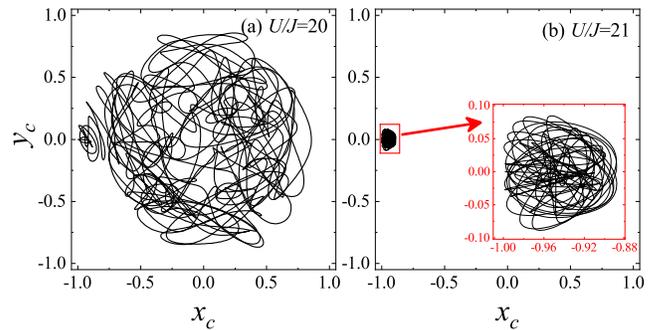}
\caption{(color online) Complex dynamic cyclotron orbit with atomic interaction. (a) $U/J=20$, (b) $U/J=21$. The other parameters are $\phi=0.5\pi$ and $K/J=1$. The inset in (b) is enlarged for more clear display.}
\label{fig:nonlinearorbit}
\end{figure}

\begin{figure}[tb]
\centering
\includegraphics[width=\columnwidth]{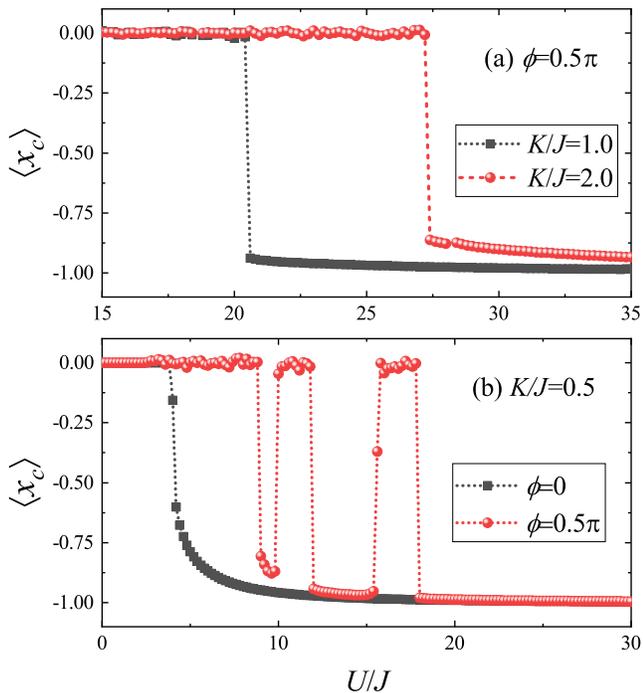}
\caption{(color online) The average of $x_c$ versus the rescaled interaction strength $U/J$: (a) for different $K/J$ at $\phi=0.5\pi$, (b) for different $\phi$ at $K/J=0.5$.  }
\label{fig:average}
\end{figure}

\begin{figure}[tb]
\centering
\includegraphics[width=\columnwidth]{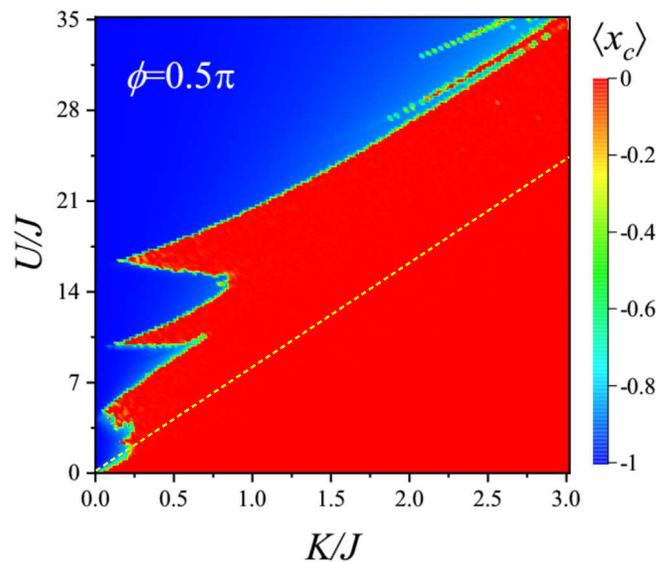}
\caption{(color online) Phase diagram of transition to self-trapping with rescaled tunneling amplitude $K/J$ and rescaled interaction strength $U/J$ at $\phi=0.5\pi$. The dashed line indicates the simple analytical results from Eq.\ref{eq:critical} with considering a double-well potential approximation. }
\label{fig:phase1}
\end{figure}

\begin{figure}[tb]
\centering
\includegraphics[width=\columnwidth]{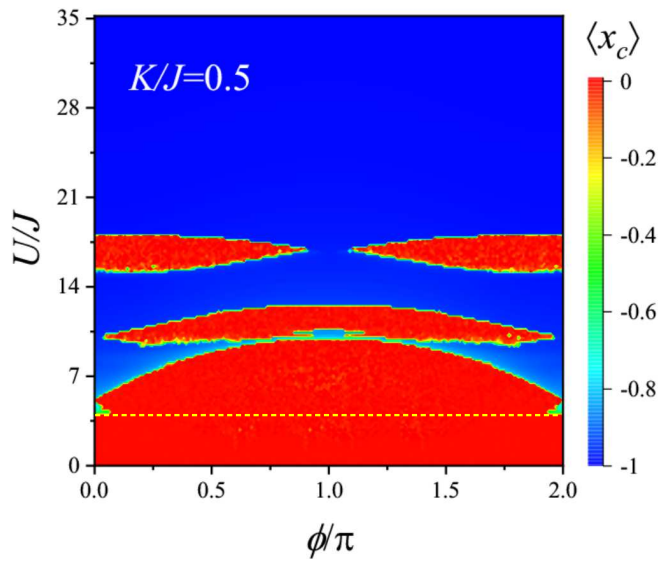}
\caption{(color online) Phase diagram of transition to self-trapping with synthetic gauge fields $\phi$ and rescaled atomic interaction strength $U/J$ at $K/J=0.5$. The dashed line indicates the simple analytical results from Eq.\ref{eq:critical} with considering a double-well potential approximation. }
\label{fig:phase2}
\end{figure}

In the presence of atomic interaction, the system (\ref{gp}) is no longer analytically solvable. It is found that the dynamics of a BEC can be strongly modified by the nonlinear atomic interaction \cite{PhysRevA.66.023404,PhysRevA.77.013402,PhysRevA.73.013619,Liu_2008,PhysRevA.78.013618}. Our numerical simulations for the effects of atomic interaction, tunneling amplitude, and synthetic gauge fields on the cyclotron dynamics have been displayed in Figs. \ref{fig:nonlinearorbit}-\ref{fig:phase2}.

When considering the atomic interaction, the system presents a rather complex dynamical cyclotron orbit. In Fig. \ref{fig:nonlinearorbit}, we show dynamic cyclotron orbit for atomic interactions $U/J=20$ and $U/J=21$, respectively. For $\phi=0.5\pi$, numerous numerical experiments show that there is a critical atomic interaction $U_c$ for certain rescaled tunneling amplitude $K/J$ and synthetic gauge fields $\phi$. When atomic interaction is less than the critical value, although the cyclotron orbit is very complicated, the mean atom position is expanded in the phase space. However, when atomic interaction is greater than the critical value, the complex cyclotron orbits exhibit obvious local characteristics in the phase space. This means that the self-trapping appears. It is known that the coupling of two macroscopic quantum states through a tunnel barrier gives rise to Josephson physics. When the interaction energy is stronger enough than the coupling energy, the transfer of particles from one macroscopic quantum state to the other is quenched, and most of the particles remain localized in one of the macroscopic quantum states. This out of equilibrium is called self-trapping \cite{PhysRevA.55.4318,PhysRevA.59.620}, which has been observed for BEC in a double-well potential \cite{PhysRevLett.79.4950}.

In order to more clearly show the effects of atomic interaction and synthetic gauge fields on the transition to self-trapping, we show the average of $x_c$ versus with the rescaled interaction strength $U/J$ at $\phi=0.5\pi$ in Fig. \ref{fig:average}(a). Obviously, there is a critical atomic interaction $U_c/J$ for certain rescaled tunneling amplitude $K/J$ and synthetic gauge fields $\phi$ in Fig. \ref{fig:average}(a). When $U/J$ is less than the critical value, the average value $\langle x_c\rangle$ is near zero, and the mean atom position is expanded in the phase space. However, when $U/J$ is greater than the critical value, the average value $\langle x_c\rangle$ is near $-1$, and then self-trapping occurs.

In Fig. \ref{fig:average}(b), we show the influences of the synthetic gauge fields on the average mean atom position $\langle x_c\rangle$. When the synthetic gauge fields are not considered, there is only a critical atomic interaction $U_c/J$ for certain rescaled tunneling amplitude $K/J$. The average value $\langle x_c\rangle$ is near zero when atomic interaction is less than the critical value. The average value $\langle x_c\rangle$ is near -1 when atomic interaction is greater than the critical value. However, considering the magnetic field, there will be several windows for the transition to self-trapping. Obviously, the synthetic gauge magnetic field exerts a significant effect on the self-trapping.

The critical behaviors of the transition to self-trapping depend on atomic interaction, tunneling amplitude, and synthetic gauge magnetic fields. In Fig. \ref{fig:phase1}, we show a phase diagram of the transition to self-trapping with rescaled tunneling amplitude $K/J$ and rescaled atomic interaction strength $U/J$ at $\phi=0.5\pi$. We can clearly see the parameter boundary of transition to self-trapping. It is a very interesting fact that the critical atomic interaction of transition to self-trapping is non-monotonic with the change of rescaled tunneling amplitude $K/J$. That is, for some rescaled tunneling amplitude $K/J$, as atomic interaction strength increases, there will be several discrete windows of self-trapping occurs.

In Fig. \ref{fig:phase2}, we show the effect of synthetic gauge fields $\phi$ and rescaled atomic interaction strength $U/J$ on the transition to self-trapping. We can see that for a certain rescaled tunneling amplitude $K/J$, when the rescaled atomic interaction strength $U/J$ is large enough, the self-trapping always occurs regardless of the change in the gauge field. Whereas, in the relatively weak atomic interaction strength regime, the critical atomic interaction of transition to self-trapping is periodically modulated by synthetic gauge fields with a period of $2\pi$. Compared with the absence of synthetic gauge fields, once the synthetic gauge fields are considered, there is always be several parameter windows for self-trapping. This is a significant feature of the influence of the synthetic gauge fields on the transition to self-trapping.

In the above discussion, we chose a particular initial state. In fact, for any initial state, the transition to self-trapping occurs at some interaction parameter, and the critical behavior in a double-well potential has been studied in our previous work \cite{PhysRevA.74.063614}.
We now look into a simple analytical expression with considering a double-well potential approximation in this system. Taking into account the above numerical results, when self-trapping occurs, the particles are all trapped in the left two wells. Therefore, we only focus on the two sets of double-well potentials in the $x$-direction. Let us express $\psi_{j}$ as $\psi_{j}=\sqrt{n_j}e^{i\theta_j}$, where the particle numbers $n_j=|\psi_{j}|^2$ and phases $\theta_j$ are all time-dependent in general. Then we further introduce the population difference $s_u=(n_2-n_1)/(n_2+n_1)$ and $s_d=(n_3-n_4)/(n_3+n_4)$, and the relative phase $\theta_u=\theta_2-\theta_1$ and $\theta_d=\theta_3-\theta_4$, using the constraint $\sum_{j} n_j=1$, we can get the classical Hamiltonian,
\begin{eqnarray}\label{classical_Hamiltonian}
H&=&-\frac{K}{2}\sqrt{1-s_u^2}\cos(\theta_u+\phi/2)+\frac{U}{16}s_u^2\nonumber\\
     && -\frac{K}{2}\sqrt{1-s_d^2}\cos(\theta_d+\phi/2)+\frac{U}{16}s_d^2.
\end{eqnarray}
Following the idea of  our previous work \cite{PhysRevA.74.063614}, the critical point is expressed as
\begin{eqnarray}\label{eq:critical}
\left(\frac{U}{K}\right)_{cr}=8\left(1+\sqrt{1-s_{u,i}^2}\cos(\theta_{u,i}+\phi/2)\right)/s_{u,i}^2.
\end{eqnarray}
In the above critical relation, $s_{u,i},$, and $\theta_{u,i}$ denoted any initial state. From the above approximate analytic expression, we see that an initial state with smaller population difference requires stronger nonlinearity so that self-trapping occurs. Moreover, the critical point can be adjusted by the relative phase between the two sets of double-well potentials and synthetic gauge fields. For example, for the case of the initial state what we considered in the above numerical results, the critical point approximates to $U/K=8$. For comparison purposes, in Figs. \ref{fig:phase1} and \ref{fig:phase2}, the dashed line indicates the simple analytical expression with considering a double-well potential approximation. It is worth emphasizing that since the system of quadruple-well potential with synthetic gauge fields is non-integrable, exact analytical results of critical behaviors are difficult to be obtained.

\section{Discussion}
In summary, we investigate the cyclotron dynamics of a BEC in a quadruple-well potential with synthetic gauge fields. We use laser-assisted tunneling to generate large tunable effective synthetic gauge fields. The mean atom position of a BEC follows an orbit that is a quantum manifestation of the classical cyclotron orbits of charged particles in a magnetic field. Without considering the atomic interaction, the mean atom position of a BEC shows periodic and quasi-periodic cyclotron orbits. We analytically give the conditions for the occurrence of periodic and quasi-periodic cyclotron orbits. Considering atomic interaction, the system exhibits transition to self-trapping. Due to the competition between synthetic gauge fields and atomic interaction, there are several discontinuous parameter windows for the transition to self-trapping, which is obviously different from that without synthetic gauge fields.

Experimentally, the quadruple-well potential can be implemented by  two possible ways. One using optical potentials \cite{PhysRevLett.107.255301,PhysRevLett.111.185301,2013Experimental}, the other using magnetic traps. Although these two methods are quite different in experimental implementation, our results are consistent for these two experimental methods \cite{PhysRevLett.106.025302}. Cyclotron orbits of the average particle position have been observed experimentally for measurement time about 13ms \cite{PhysRevLett.107.255301}. During the measurement time, more than four cycles of cyclotron orbits were observed.  In previously reported realizations of self-trapping of BEC in double-well potentials \cite{PhysRevLett.95.010402} the time scale can reach on the order of 50 ms. Our theory predicts the periodic and quasi-periodic cyclotron orbits as well as transition to self-trapping of a BEC with synthetic gauge fields. We hope our discussion
will stimulate further experiments.

\section{Acknowledgments}
This work is supported by the National Natural Science Foundation of China (Grant No. 12005173), by the Natural Science Foundation of Gansu Province (Grant No. 20JR10RA082), by the China Postdoctoral Science Foundation (Grant No. 2020M680318), and by the NSAF (Grant No. U1930402 and No. U1930403).
\bibliography{Cyclotron-dynamics}


\end{document}